# Frequency modulation of THz emission from a laser filament array

Sean D. McGuire[1,*] and Mikhail N. Shneider[2]

[1]Université Paris Saclay, Gif-sur-Yvette 91190, France
[2]Princeton University, Princeton, New Jersey 08544, USA

**ABSTRACT**. A simple traveling wave antenna model is used to theoretically study emission from an array of filaments. Both transverse and longitudinal arrays of filaments are considered. The angular distribution and power in the THz signal are significantly modified, which is consistent with other trends reported in the literature. The frequency content of the THz emission signal is also strongly modified under certain conditions. Whereas the emission from a single filament is broadband, the emission from a periodic transverse array consists of several discrete frequencies. For this latter case, the THz spectrum can be approximated as the product of two factors – the spectrum from a single filament and a complex frequency-dependent phase factor associated with the spatial distribution of the filaments. The complex phase factor accounts for interference effects, amplifying certain frequencies from the single filament spectrum while suppressing others. The location and width of the discrete frequencies present in the final spectrum depends on the number of filaments and their spacing. These results point to an interesting possibility of tailoring the frequency content in the THz signal.

## I. INTRODUCTION.

Tailoring the frequency content of THz emission has been a topic of interest in the literature. One way to generate THz emission is by focusing femtosecond pulses onto nonlinear crystals, and several authors have looked at methods of controlling the frequency of this emission [1-5]. For example, D'Amico *et al* [1] and Vidal *et al* [2, 3] show that optical pulse shaping of the femtosecond laser can be used to produce narrowband tunable THz emission. Other work has shown that such pulse shaping can also be used to control the THz frequency via the 2-color mechanism in gas filaments [6, 7]. Fundamentally, these techniques rely on the interference between several THz pulses that are initiated in the active medium. Interference effects have been observed to occur in arrays of gas filaments as well, both when using the 1-color [8-13] and 2-color [14, 15] emission mechanisms. Such effects have been shown to significantly enhance the THz emission signal and alter the angular emission distribution relative to what is expected from a single filament. Furthermore, spectral alterations due to interference effects in a 2-filament array have also been experimentally demonstrated [16-18].

We present a simple model capable of capturing interference effects in a filament array, which is based around the well-known Traveling Wave Antenna (TWA) model for the 1-color Transition-Cherenkov mechanism illustrated in FIG 1 [19, 20]. The calculations presented show that the spectrum of the THz signal from a filament array can be approximated as the product of two factors – the spectrum from a single filament and a complex frequency-dependent amplitude and phase factor associated with the spatial distribution of the filaments. This factor accounts for interference effects and leads to significant alterations in the spectral content of the emission. Such alterations have been experimentally demonstrated for a 2-filament array [16, 17]. However, the calculations suggest that moving from a 2-filament to a multiple-filament array further accentuates this spectral modulation. For example, in the case of a transverse multiple-filament array considered below in Section IIIA, the spectral content consists of several narrowband emission peaks. Such a transverse filament array could be generated using a microlens array or periodic mesh [21]. Recent work has shown that a slight modification of the underlying TWA model used here enables it to simulate emission due to the 2-color mechanism [22]. Therefore, though the 2-color is not considered in this paper, the conclusions drawn may also have implications for this method of THz generation as well.

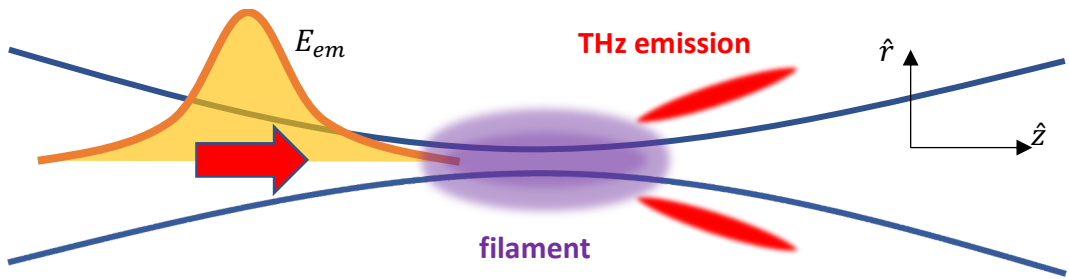

FIG 1: Experimental configuration used for the generation of THz radiation from a single filament. The direction of the emitted THz radiation depends upon the experimental parameters used but generally falls between the forward ($+\hat{z}$) and side directions.

*Contact author: sean.mc-guire@centralesupelec.fr

## II. MODEL

The 1D model used here is taken directly from Ref. [23], which itself is based upon prior work in the literature [19, 20, 24]. According to this model, the electric field emitted by a single filament is as follows:

$$\hat{E}_{SF}(r,\theta,\omega) = \frac{\mu_0 c}{4\pi r}\hat{\imath}(\omega)\frac{\sin\theta}{\xi}e^{-i\frac{\omega n}{c}r}\left[1 - e^{-i\frac{\omega L}{u}\xi}\right] \quad 1$$

where $\theta$ is the angle between the axis of laser propagation and detector, $r$ is the detector distance, $I$ is the propagating current, $\xi = 1 - \beta n \cos\theta$, $\beta = u/c$, $L$ is the length of the antenna, $u$ is the speed at which the current propagates, and $n$ is the index of refraction at the THz emission wavelength. For the calculations in this paper, it is assumed that $\beta = 1$ and $n = 1$. The subscript '$SF$' represents the fact that the emission is from a '*single filament*'. $\hat{\imath}(\omega)$ is the Fourier transform of a propagating laser-driven current packet and is a complex variable. An expression for this current can be derived that accounts for ponderomotive, radiation pressure, and ionization current sources. This expression is as follows:

$$\hat{\imath}(\omega) = \frac{\hat{\gamma}(\omega)}{i\omega + \nu_e}S \quad 2$$

with $\hat{\gamma}(\omega)$ being the Fourier transform of $\gamma(t)$, where:

$$\gamma(t) = \gamma(t)\,\hat{z} = -\frac{1}{e\varepsilon_0 c^2}\left(\frac{e^2}{m_e}\right)^2\frac{1}{\omega_L^2 + \nu_e^2} \ast \left[\frac{n_0}{2}\frac{dI}{d\eta} + \nu_e n_0 I + I\frac{dn_0}{d\eta}\right] \quad 3$$

$I(t) = (1/2)c\varepsilon_0 \bar{E}_1 \bar{E}_1^*$ is the intensity of the laser, $n_0(t)$ is the electron density, and $\omega_L$ is the carrier frequency of the laser. The above relation assumes that $\nu_e$ is constant. EQNS 2 – 3 require a certain number of inputs: the length of the antenna $L$, cross-sectional area of the antenna $S$, electron density $n_0$, and laser pulse profile $I(t)$. These can all be estimated using a separate beam propagation model which predicts the evolution of laser intensity throughout the focal zone. For this paper, the single filament current profile $\hat{\imath}(\omega)$ was taken directly from Ref. [23], which simulated the focusing of a $50\,fs$ and $800\,nm$ wavelength pulse using a $50\,cm$ focal length lens. The pulse energy was $30\,\mu J$ pulse. The resulting filament length and radius were calculated to be $L = 2.5\,mm$ and $R = 7\,\mu m$. The resulting emission is shown in FIG 2A.

The emission from a generic distribution of several filaments can be calculated by summing the electric field from the individual filaments.

$$\hat{E}(r,\theta,\omega) = \sum_{m=0}^{N}\left(\hat{E}_{SF}\right)_m \quad 4$$

In general, this is a vector sum given that the various $\left(\hat{E}_{SF}\right)_m$ are vector quantities. The sum can be calculated numerically for an arbitrary distribution of filaments. We apply a few simplifying assumptions for all calculations. First, the filaments are assumed to be identical. Second, emission from one filament is assumed to be unimpacted by all other filaments. Third, the far-field approximation is invoked – the spatial extent of the array is assumed much smaller than the distance to the detector ($r$). These assumptions significantly simplify EQN 4. Given that the filaments are identical and closely spaced relative to the detector distance, the vector electric fields $\left(\hat{E}_{SF}\right)_m$ from the individual filaments can be assumed to share the same spatial orientation and polarization. Thus EQN 4 becomes a scalar sum. Additionally, the variables $\theta$ and $r$ in EQN 1 are taken to be the same for various filaments: one important exception being within the exponential term $e^{-i\frac{\omega n}{c}r}$, where the variation in $r$ cannot be ignored and accounts for interference between the emission from various filaments. This latter variation in $r$ will be addressed below.

## III. RESULTS

### A. Transverse array of filaments

Although transverse bundles of femtosecond filaments have been shown to occur within a single focused beam [25], the 1D periodic array discussed in this section would require the use of several beams. In this scenario, $N$ filaments are organized into a 1D periodic array (spatial period $D$) orthogonal to the axes of laser propagation. The filaments are assumed to be generated at the same moment in time. Note that the emission from a single filament is axisymmetric about the axis of laser propagation. However, this geometry results in THz emission that is not axisymmetric. The 1D filament array and detector define a plane, and the emission is altered upon rotation of this plane about any one of the filament axes. In the simple analytical expressions shown below, only the in-plane emission is considered. In this case, Eqn. 4 becomes:

$$\hat{E}(r,\theta,\omega) = \frac{\mu_0 c}{4\pi r_0} \hat{\imath}(\omega) \frac{\sin\theta}{\xi}$$
$$* \left[1 - e^{-i\frac{\omega L}{u}\xi}\right] \sum_{m=0}^{N} e^{-i\frac{\omega}{c} r_m} \quad 5$$

Assuming that $ND \ll r_m$, $r_m$ can be approximated as $r_m \sim r_0 + mD \sin\theta$ (where $r_0$ is the distance to the detector from the closest filament denoted by subscript '0'). Substituting into Eqn. 5 and simplifying gives:

$$\hat{E}(r,\theta,\omega) = \frac{\mu_0 c}{4\pi r_0} \hat{\imath}(\omega) \frac{\sin\theta}{\xi} \left[1 - e^{-i\frac{\omega L}{u}\xi}\right]$$
$$* e^{-i\frac{\omega}{c} r_0} \sum_{m=0}^{N} e^{-i\frac{\omega}{c} mD \sin\theta}$$
$$= \left(\hat{E}_{SF}\right)_{r_0} \sum_{m=0}^{N} e^{-i\frac{\omega}{c} mD \sin\theta} \quad 6$$

In other words, the electric field can be seen as the electric field from a single filament but modified by a phase factor which accounts for interference effects, and which mathematically corresponds to a Fourier series with a fundamental period of $c/D \sin\theta$ in frequency ($\Delta f = \Delta\omega/2\pi$) space.

FIG 2B and FIG 2C show the emission that results from a periodic array of $N = 2$ and $N = 10$ filaments. The period between filaments was taken to be $1\,mm$. While the amplitude of the emission has significantly increased, this is to be expected as more filaments are radiating. The angular distribution of emission is not significantly impacted with respect to the emission from a single filament. However, the frequency content at the maximum angle of emission is significantly affected. Moving from a single filament to a periodic array results in a modulation of the frequency content. If more filaments are included, this effect becomes more pronounced such that the emission may be approximated as a sum of several discrete frequencies.

The observations above suggest that – given a particular $\hat{\imath}(\omega)$ – something approaching monochromatic emission can be obtained. Given the frequency bandwidth of the emission from a single filament, the filament separation can be chosen accordingly. For the case considered, if the period is decreased to $D = 300\,\mu m$, FIG 2D shows the resulting emission. The emission is now almost monochromatic – the central peak being at about $12\,THz$. Further increasing the number of filaments makes the width of each resonant emission peak narrower and suppresses emission at other frequencies.

EQN 4 can also be used to model the emission from a more generalized transverse 2D mesh of filaments. We have performed such calculations and, aside from the lack of axisymmetry, the conclusions of the simple model presented here apply. Furthermore, by using a sufficiently large square array of filaments (e.g. $10 \times 10$), the emission can be rendered approximately axisymmetric upon rotation about an axis which is normal to the 2D plane of filaments, and which passes through the center point of the mesh. Using large numbers of filaments also results in a much larger THz emission signal. For example, a $10 \times 10$ periodic lattice (period $D = 1\,mm$) of filaments results in a maximum energy of $2.7\,pJ/sr$ at an angle of $\theta = 6.5°$, versus $4.4\,fJ/sr$ at an angle of $\theta = 5°$ from a single filament (FIG 2A). This represents an increase of ~600 in the average intensity of the THz signal, for a corresponding increase in laser energy of $10^2 = 100$.

### B. Longitudinal array of filaments

The filaments in this scenario are spaced at periodic intervals along a single propagation axis – which could be the propagation of one or potentially several optical beams. This reconfiguration leads to three primary changes with respect to the case of a transverse array of filaments:

1. The distance to the detector becomes $r_m \sim r_0 + mD \cos\theta$ (where $r_0$ is the distance to the detector from the closest filament).
2. The filaments are not assumed to occur at the same moment in time. Rather, each successive filament is initiated at a time delay of $D/c$, where $c$ is the speed of light. The shift property of the Fourier transform implies that this leads to an additional phase factor for each successive filament, equal to $e^{i\omega mD/c}$. This phase factor should be applied to the current term associated with filament $m$.
3. The THz emission from the 1D array is axisymmetric about the array axis.

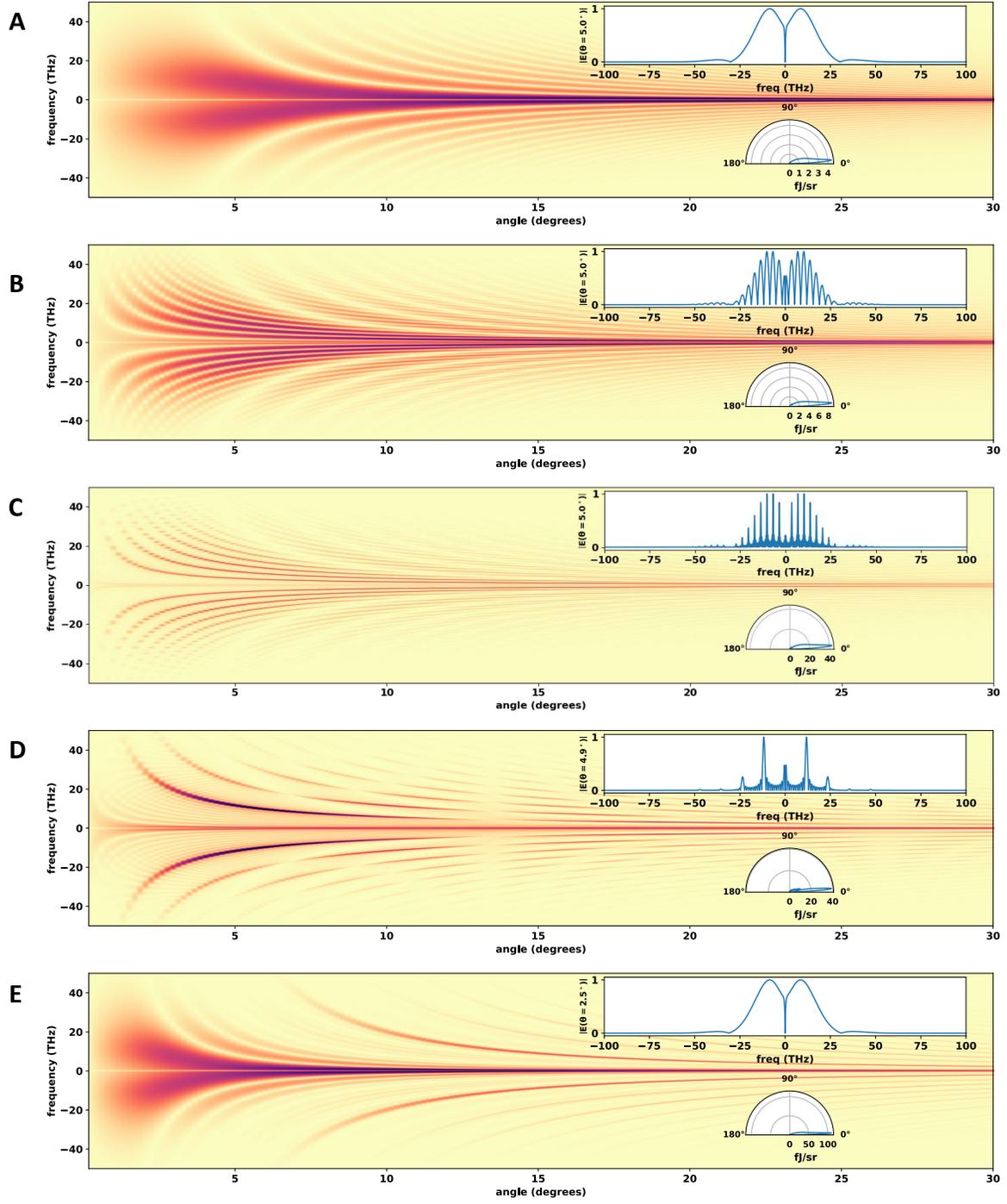

FIG 2: Contour plots of the THz emission electric field strength $\left|\hat{E}(r \to \infty, \theta, f = \omega/2\pi)\right|$ in the far-field for various filament arrangements. The spectrally integrated emission versus angle $\theta$ and the frequency content of the electric field at the angle of maximum emission ($\left|\hat{E}(r \to \infty, \theta = \theta_{max}, f = \omega/2\pi)\right|$) are shown as insets. (A) Emission from a single filament. (B) Transverse array of 2 filaments, $D = 1\ mm$. (C) Transverse array of 10 filaments, $D = 1\ mm$. (D) Transverse array of 10 filaments, $D = 300\ \mu m$. (E) Longitudinal array of 10 filaments, $D = 1\ mm$.

These changes lead to the following equation:

$$\hat{E}(r,\theta,\omega) = \frac{\mu_0 c}{4\pi r_0}\hat{i}_0(\omega)\frac{\sin\theta}{\xi}\left[1 - e^{-i\frac{\omega L}{u}\xi}\right]$$
$$* e^{-i\frac{\omega}{c}r_0}\sum_{m=0}^{N} e^{i\frac{\omega}{c}mD(1-\cos\theta)}$$
$$= \left(\hat{E}_{SF}\right)_{r_0}\sum_{m=0}^{N} e^{i\frac{\omega}{c}mD(1-\cos\theta)} \qquad 7$$

where $\hat{i}_0(\omega)$ should be understood to represent the current profile of the first filament (which is shifted in time from the current profile of the other filaments).

FIG 2C shows the angular distribution and frequency content for $N = 10$ filaments with a periodic filament spacing of $D = 1\,mm$. Here the frequency content at the maximum angle of emission if only slightly effected. Meanwhile, the angular distribution is affected in the sense that the emission becomes increasingly oriented towards the forward direction as $N$ is increased.

## V. CONCLUSIONS

THz emission from a periodic array of filaments is examined using a simplified TWA antenna formalism. The model developed here was used to study transverse and longitudinal periodic arrays. The input current profile $\hat{i}(\omega)$, filament length $L$, filament spacing $D$, and number of filaments $N$ all play critical roles in the redistribution and frequency content of the emission. The transverse periodic array appears of particular interest as the frequency content in the signal can be significantly modified, leading to the emission of discrete frequencies and possibly quasi-monochromatic emission. Similar interference effects have been experimentally observed for a 2-filament array [16], which suggests that the calculations shown here should remain valid for a multiple-filament array. Such an approach for tailoring the frequency content of the THz signal provides an alternative to existing approaches involving, for example, the use of nonlinear crystals [1-5] or metamaterials [26].

## ACKNOWLEDGMENTS
This work was partially supported by the Princeton Collaborative Low Temperature Plasma Research Facility (PCRF).